\documentclass[adp,a4paper,fleqn%
]{w-art}
\usepackage{times,cite,w-thm}

\begin{document}
\DOIsuffix{xxx}
\Volume{xx}
\Month{xx}
\Year{20xx}
\pagespan{1}{}
\Receiveddate{XXXX}
\Reviseddate{XXXX}
\Accepteddate{XXXX}
\Dateposted{XXXX}

\title[Covariant dynamics on the brane]{3+1+1 dimensional covariant gravitational dynamics on an asymmetrically embedded brane}

\author[Z. Keresztes]{Zolt\'{a}n Keresztes\inst{1,2}}
\author[L.\'A. Gergely]{L\'{a}szl\'{o} \'{A}. Gergely\inst{1,2,3}} 
\address[\inst{1}]{Department of Theoretical Physics, University of Szeged, Tisza Lajos krt 84-86, Szeged 6720, Hungary}
\address[\inst{2}]{Department of Experimental Physics, University of Szeged, D\'{o}m t\'{e}r 9, Szeged 6720, Hungary}
\address[\inst{3}]{Institute for Advanced Study, Collegium Budapest, Szenth\'aroms\'ag u 2, Budapest 1014, Hungary}

\begin{abstract}
We give the evolution and constraint equations on an asymmetrically embedded brane in the form of average and difference equations.  
\end{abstract}

\maketitle

\section{Introduction}

In a recent paper \cite{3+1+1} we have developed the covariant gravitational
dynamics in a 3+1+1 dimensional space-time in the spirit of the general
relativistic 3+1 covariant cosmology \cite{3+1}, generalizing both previous
approaches to 5-dimensional (5d) gravitational dynamics \cite{brane-dynamics}%
, brane 3+1 covariant cosmology \cite{Maartens-eqs} and 2+1+1 covariant
dynamics \cite{2+1+1}. Such a formalism may turn useful in discussing
perturbations on the brane \cite{branepert}. The singled-out directions are
the normal $n^{a}$ to the hypersurface representing the time-evolving
3-dimensional space (the brane with metric $h_{ab}$ and tension $\lambda $)
and a temporal direction $u^{a}$ tangent to this hypersurface. All employed
gravitational variables are projected to the brane. They consist of
kinematical variables $(\Theta ,\sigma _{ab},\omega _{ab},A_{a},K_{a},L_{a})$
related to the vector $u^{a}$; analogous quantities, carrying an overhat,
related to $n^{a}$; two kinematical scalars $(K,\hat{K})$ related to both;
finally gravito-electro-magnetic quantities $(\mathcal{E},\mathcal{H}_{k},%
\mathcal{F}_{kl},\mathcal{E}_{k},\widehat{\mathcal{E}}_{k},\mathcal{E}_{kl},%
\widehat{\mathcal{E}}_{kl},\mathcal{H}_{kl},\widehat{\mathcal{H}}_{kl})$.
For details on the definitions of these quantities see Ref. \cite{3+1+1}.
The matter on the brane has been decomposed as $T_{ab}=\rho
u_{a}u_{b}\!+\!q_{(a}u_{b)}\!+\!ph_{ab}\!+\!\pi _{ab},$ while possible
non-standard model fields nesting in the 5d space-\negthinspace time as $%
\widetilde{T}_{ab}=\widetilde{\rho }u_{a}u_{b}\!+\!2\widetilde{q}%
_{(a}u_{b)}\!+\!2\widetilde{q}u_{(a}n_{b)}\!+\!\widetilde{p}h_{ab}\!+\!%
\widetilde{\pi }n_{a}n_{b}\!+\!2\widetilde{\pi }_{(a}n_{b)}\!+\!\widetilde{%
\pi }_{ab}.$ The gravitational coupling constants $\widetilde\kappa^2$ and $%
\kappa^2=\widetilde\kappa^4\lambda/6$ act in 5d and on the brane,
respectively.

The generic form of the evolution and constraint equations in the
5-dimensional spacetime were given as Appendix C of Ref. \cite{3+1+1} and
they were specified on a $Z_{2}$-symmetrically embedded brane in Subsection
IV.D. The embedding however is not necessarely symmetric: an asymmetric
embedding is known to generate late time acceleration in a cosmological
setup \cite{Induced}. Therefore here we generalize the formalism to an
asymmetrically embedded brane. Following the recipe presented in Subsection
IV.D of Ref. \cite{3+1+1}, we give here the evolution and constraint
equations obtained both as averages and differences across the brane. We
denote the average over the two sides of the brane of any quantity by angle
brackets and the jump by $\Delta $. For the extrinsic curvature components $%
\mathcal{K}\equiv (\widehat{\Theta },\widehat{\sigma }_{ab},\widehat{K},%
\widehat{K}_{a})$ the latter is directly related to the brane matter
variables and brane tension cf. the Israel-Lanczos condition, Eqs. (76)-(79)
of Ref. \cite{3+1+1}. Angle brackets on indices indicate projection to the
3-space, symmetrization and trace-free character.

\section{The average equations}

The evolution equations arising as averages are:%
\[
0=\langle \dot{\widehat{\Theta }}\rangle \!-\!D^{a}\langle \widehat{K}%
_{a}\rangle \!+\!\langle \widehat{K}\!+\!\frac{\widehat{\Theta }}{3}\rangle
\Theta \!-\!2\langle \widehat{K}^{a}\rangle A_{a}\!+\!\langle \widehat{%
\sigma }_{ab}\rangle \sigma ^{ab}\!-\!\widetilde{\kappa }^{2}\langle 
\widetilde{q}\rangle \ ,
\]%
\[
0=\langle \dot{\widehat{K}}_{\langle a\rangle }\rangle \!-\!D_{a}\langle 
\widehat{K}\!-\!\frac{2}{3}\widehat{\Theta }\rangle \!-\!D^{b}\langle 
\widehat{\sigma }_{ab}\rangle \!+\!\frac{4\Theta }{3}\langle \widehat{K}%
_{a}\rangle \!-\!\langle \widehat{K}\!+\!\frac{\widehat{\Theta }}{3}\rangle
A_{a}\!-\!\langle \widehat{\sigma }_{ab}\rangle A^{b}\!-\!\omega
_{ab}\langle \widehat{K}^{b}\rangle \!+\!\sigma _{ab}\langle \widehat{K}%
^{b}\rangle \!+\!\widetilde{\kappa }^{2}\langle \widetilde{\pi }_{a}\rangle
\ ,
\]%
\begin{gather*}
0=\dot{H}_{\langle kj\rangle }\!+\!\varepsilon _{ab\langle
k}D^{a}E_{j\rangle }^{b}\!+\!\frac{1}{2}\varepsilon _{ab\langle
k}D^{a}\langle \widehat{\mathcal{E}}_{j\rangle }^{b}\rangle \!+\!\Theta
H_{kj}\!-\!3\sigma _{a\langle k}H_{j\rangle }^{a}\!-\!\omega _{a\langle
k}H_{j\rangle }^{a}\!-\!2\varepsilon _{\langle k}^{\ \ \ ab}E_{j\rangle
a}A_{b}\! \\
-\frac{3}{2}\langle \widehat{\mathcal{E}}_{\langle j}\rangle \omega
_{k\rangle }-\!\frac{1}{2}\varepsilon _{\langle k}^{\ \ \ ab}\sigma
_{j\rangle a}\langle \widehat{\mathcal{E}}_{b}\rangle \!-\!\frac{1}{2}%
\varepsilon _{\langle k}^{\ \ \ \ cd}\langle \widehat{\sigma }_{j\rangle
c}\rangle D_{d}\langle \widehat{K}\!-\!\frac{\widehat{\Theta }}{3}\rangle
\!+\!\frac{1}{2}\langle \widehat{K}\!-\!\frac{\widehat{\Theta }}{3}\rangle
\varepsilon _{ab\langle k}D^{a}\langle \widehat{\sigma }_{j\rangle
}^{b}\rangle \! \\
+\frac{\langle \widehat{\sigma }^{cb}\rangle \!}{2}\varepsilon _{ab\langle
k}D^{a}\langle \widehat{\sigma }_{j\rangle c}\rangle +\!\langle \widehat{%
\Theta }\rangle \langle \widehat{K}_{\langle k}\rangle \omega _{j\rangle }+\!%
\frac{1}{2}\varepsilon _{ab\langle k}\langle \widehat{\sigma }_{j\rangle
c}\rangle D^{a}\langle \widehat{\sigma }^{cb}\rangle \!+\!\frac{\langle 
\widehat{\Theta }\rangle }{3}\varepsilon _{\langle k}^{\ \ \ ab}\sigma
_{j\rangle a}\langle \widehat{K}_{b}\rangle \! \\
-\frac{1}{2}\varepsilon _{ab\langle k}\langle \widehat{K}_{j\rangle }\rangle
D^{a}\langle \widehat{K}^{b}\rangle \!-\!\frac{1}{2}\varepsilon _{ab\langle
k}\langle \widehat{K}^{b}\rangle D^{a}\langle \widehat{K}_{j\rangle }\rangle
+\!\frac{\langle \widehat{K}_{c}\rangle \!}{2}\varepsilon _{\langle k}^{\ \
\ ab}\sigma _{j\rangle b}\langle \widehat{\sigma }_{a}^{c}\rangle -\!\frac{%
3\langle \widehat{K}_{a}\rangle }{2}\langle \widehat{\sigma }_{\langle
j}^{a}\rangle \omega _{k\rangle }\! \\
\!+\frac{\widetilde{\kappa }^{4}}{8}\!\biggl[\varepsilon _{ab\langle
k}D^{a}\pi _{j\rangle c}\pi ^{cb}\!+\!\varepsilon _{ab\langle k}\pi
_{j\rangle c}D^{a}\pi ^{cb}-\!\frac{1}{3}\varepsilon _{\langle k}^{\ \ \ \
cd}\pi _{j\rangle c}D_{d}\left( \rho \!+\!3p\right) \!-\!\varepsilon
_{ab\langle k}q^{b}D^{a}q_{j\rangle }\! \\
-\frac{\left( 2\lambda \!-\!\rho \!-\!3p\right) }{3}\varepsilon _{ab\langle
k}D^{a}\pi _{j\rangle }^{b}\!-\!\left( \lambda \!+\!\rho \right) \omega
_{\langle k}q_{j\rangle }\!-\!\frac{2\left( \lambda \!+\!\rho \right) }{3}%
\varepsilon _{\langle k}^{\ \ \ ab}\sigma _{j\rangle a}q_{b}\!-\!\varepsilon
_{ab\langle k}q_{j\rangle }D^{a}q^{b} \\
\!-\varepsilon _{\langle k}^{\ \ \ ab}\sigma _{j\rangle b}\pi
_{a}^{c}q_{c}\!+\!3\pi _{\langle j}^{a}\omega _{k\rangle }q_{a}\!\biggr]\!-\!%
\frac{\widetilde{\kappa }^{2}}{3}\varepsilon _{ab\langle k}D^{a}\langle 
\widetilde{\pi }_{j\rangle }^{b}\rangle \!-\!\widetilde{\kappa }^{2}\langle 
\widetilde{q}_{\langle j}\rangle \omega _{k\rangle }\!-\!\frac{\widetilde{%
\kappa }^{2}}{3}\varepsilon _{\langle k}^{\ \ \ \ ab}\sigma _{j\rangle
a}\langle \widetilde{q}_{b}\rangle \ ,
\end{gather*}%
\begin{gather*}
0=\dot{E}_{\langle kj\rangle }\!-\!\frac{1}{2}\langle \dot{\widehat{\mathcal{%
E}}}_{\langle kj\rangle }\rangle \!-\!\varepsilon _{ab\langle
k}D^{a}H_{j\rangle }^{b}\!+\!\frac{1}{2}D_{\langle k}\langle \widehat{%
\mathcal{E}}_{j\rangle }\rangle \!+\!\Theta E_{kj}\!-\!\frac{\Theta }{6}%
\langle \widehat{\mathcal{E}}_{kj}\rangle \!+\!\langle \widehat{\mathcal{E}}%
_{\langle k}\rangle A_{j\rangle }\!-\!\frac{2}{3}\langle \mathcal{E}\rangle
\sigma _{kj} \\
\!-\frac{\langle \widehat{\mathcal{E}}_{\langle j}^{a}\rangle }{2}\left(
\omega _{k\rangle a}\!+\!\sigma _{k\rangle a}\right) \!+\!E_{\langle
k}^{a}\left( \omega _{j\rangle a}\!-\!3\sigma _{j\rangle a}\right)
\!+\!2\varepsilon _{\langle k}^{\ \ \ \ ab}H_{j\rangle a}A_{b}\!-\!\langle 
\widehat{\sigma }_{\langle j}^{a}\rangle \langle \dot{\widehat{\sigma }}%
_{k\rangle a}\rangle \!-\!\frac{\langle \widehat{\sigma }_{kj}\rangle }{2}%
\langle \widehat{K}\!-\!\frac{\widehat{\Theta }}{3}\rangle ^{\cdot } \\
\!+\langle \widehat{K}_{\langle k}\rangle D_{j\rangle }\langle \widehat{K}%
\!-\!\widehat{\Theta }\rangle \!-\!\frac{\langle \widehat{\Theta }\rangle }{3%
}D_{\langle k}\langle \widehat{K}_{j\rangle }\rangle \!+\!\frac{\langle 
\widehat{\sigma }_{\langle j}^{a}\rangle }{2}D_{k\rangle }\langle \widehat{K}%
_{a}\rangle \!+\!\frac{\langle \widehat{K}_{a}\rangle }{2}D_{\langle
k}\langle \widehat{\sigma }_{j\rangle }^{a}\rangle +\!\frac{\langle \widehat{%
\Theta }\rangle }{3}\langle \widehat{K}\!+\!\frac{\widehat{\Theta }}{3}%
\rangle \sigma _{kj} \\
\!+\langle \widehat{K}_{\langle k}\rangle \!\left[ D^{b}\langle \widehat{%
\sigma }_{j\rangle b}\rangle \!+\!\frac{3}{2}\omega _{j\rangle a}\langle 
\widehat{K}^{a}\rangle \!+\!\langle \widehat{\sigma }_{j\rangle a}\rangle
A^{a}\!-\!\frac{1}{2}\sigma _{j\rangle b}\langle \widehat{K}^{b}\rangle \!-\!%
\frac{7\Theta }{6}\langle \widehat{K}_{j\rangle }\rangle \right] \!\!-\!%
\frac{\langle \widehat{\sigma }_{a}^{c}\rangle }{2}\langle \widehat{\sigma }%
_{\langle j}^{a}\rangle \omega _{k\rangle c} \\
\!-\!\frac{1}{2}\langle \widehat{K}\!-\!\frac{\widehat{\Theta }}{3}\rangle \!%
\left[ \langle \dot{\widehat{\sigma }}_{\langle kj\rangle }\rangle \!+\!%
\frac{\Theta }{3}\langle \widehat{\sigma }_{kj}\rangle \!-\!2\langle 
\widehat{K}_{\langle k}\rangle A_{j\rangle }\!+\!\langle \widehat{\sigma }%
_{\langle k}^{a}\rangle \left( \omega _{j\rangle a}\!+\!\sigma _{j\rangle
a}\right) \right] \!\!-\!\frac{\sigma _{jk}}{2}\langle \widehat{K}%
^{a}\rangle \langle \widehat{K}_{a}\rangle \! \\
\!+\langle \widehat{\sigma }_{a\langle j}\rangle A_{k\rangle }\langle 
\widehat{K}^{a}\rangle \!-\!\frac{\langle \widehat{\sigma }_{c}^{a}\rangle }{%
2}\langle \widehat{\sigma }_{\langle k}^{c}\rangle \sigma _{j\rangle a}\!-\!%
\frac{\Theta }{6}\langle \widehat{\sigma }_{\langle j}^{a}\rangle \langle 
\widehat{\sigma }_{k\rangle a}\rangle +\!\frac{\widetilde{\kappa }^{4}}{8}\!%
\biggl[\!\frac{2}{3}q_{\langle k}D_{j\rangle }\left( \rho \!-\!3p\right)
\!-\!\frac{\left( \rho \!+\!3p\right) ^{\cdot }}{3}\pi _{kj}\! \\
\!\!-2\pi _{\langle j}^{a}\dot{\pi}_{k\rangle a}+\!\frac{2\left( \lambda
\!+\!\rho \right) }{3}D_{\langle k}q_{j\rangle }\!-\!\pi _{\langle
j}^{a}D_{k\rangle }q_{a}-\!q_{a}D_{\langle k}\pi _{j\rangle }^{a}\!+\!\frac{%
2\left( \lambda \!+\!\rho \right) }{3}\left( \rho \!+\!p\right) \sigma
_{kj}\!-\!\pi _{\langle j}^{a}\omega _{k\rangle c}\pi _{a}^{c} \\
-\sigma _{jk}q^{a}q_{a}\!-\!2\pi _{a\langle j}A_{k\rangle
}q^{a}+\!q_{\langle k}\!\!\left( \!-\!2D^{b}\pi _{j\rangle b}\!+\!3\omega
_{j\rangle a}q^{a}\!-\!2\pi _{j\rangle a}A^{a}\!-\!\sigma _{j\rangle
b}q^{b}\!-\!\frac{7\Theta }{3}q_{j\rangle }\right) \!-\!\pi _{c}^{a}\pi
_{\langle k}^{c}\sigma _{j\rangle a} \\
\!+\frac{\left( 2\lambda \!-\!\rho \!-\!3p\right) }{3}\!\left( \dot{\pi}%
_{\langle kj\rangle }\!+\!\frac{\Theta }{3}\pi _{kj}\!+\!2q_{\langle
k}A_{j\rangle }\!+\!\pi _{\langle k}^{a}\omega _{j\rangle a}\!+\!\pi
_{\langle k}^{a}\!\sigma _{j\rangle a}\right) \!-\!\frac{\Theta }{3}\pi
_{\langle j}^{a}\pi _{k\rangle a}\!\biggr]\!-\!\frac{\widetilde{\kappa }^{2}%
}{3}\!\biggl[\langle \dot{\widetilde{\pi }}_{\langle kj\rangle }\rangle  \\
\!\!\!\!+\!D_{\langle k}\langle \widetilde{q}_{j\rangle }\rangle
\!\!-\!\langle \widetilde{\pi }_{\langle k}\rangle \langle \widehat{K}%
_{j\rangle }\rangle \!\!-\!\frac{\widetilde{\kappa }^{2}}{4}q_{\langle
k}\!\Delta \widetilde{\pi }_{j\rangle }\!+\!4\langle \widetilde{q}_{\langle
k}\rangle A_{j\rangle }\!\!+\!\langle \widetilde{\rho }\!+\!\widetilde{p}%
\rangle \sigma _{jk}\!+\!\frac{\Theta }{3}\langle \widetilde{\pi }%
_{jk}\rangle \!+\!\langle \widetilde{\pi }_{\langle j}^{a}\rangle \!\!\left(
\omega _{k\rangle a}\!+\!\sigma _{k\rangle a}\right) \!\biggr]\ ,
\end{gather*}%
\begin{gather*}
0=\langle \dot{\widehat{\mathcal{E}}}_{\langle k\rangle }\rangle \!+\!\frac{%
4\Theta }{3}\langle \widehat{\mathcal{E}}_{k}\rangle \!-\!\frac{1}{3}%
D_{k}\langle \mathcal{E}\rangle \!-\!\frac{4\langle \mathcal{E}\rangle }{3}%
A_{k}\!-\!D^{a}\langle \widehat{\mathcal{E}}_{ka}\rangle \!-\!\langle 
\widehat{\mathcal{E}}_{ka}\rangle A^{a}\!-\!\left( \omega _{ka}\!-\!\sigma
_{ka}\right) \langle \widehat{\mathcal{E}}^{a}\rangle \!+\!\langle \widehat{K%
}^{a}\rangle \! \\
\times \biggl[\langle \dot{\widehat{\sigma }}_{\langle ka\rangle }\rangle
+\!\sigma _{ck}\langle \widehat{\sigma }_{a}^{c}\rangle \!-\!2D_{k}\langle 
\widehat{K}_{a}\rangle \!+\!D_{a}\langle \widehat{K}_{k}\rangle
\!-\!2A_{\langle k}\langle \widehat{K}_{a\rangle }\rangle \!-\!\sigma
_{ba}\langle \widehat{\sigma }_{k}^{b}\rangle \!+\!\varepsilon _{cab}\omega
^{b}\langle \widehat{\sigma }_{k}^{c}\rangle \!\biggr]\!+\!\langle \widehat{%
\sigma }^{ab}\rangle \! \\
\times \biggl[D_{k}\langle \widehat{\sigma }_{ab}\rangle \!+\!\frac{2}{3}%
\langle \widehat{K}_{k}\rangle \sigma _{ab}\!\biggr]\!+\!\langle \widehat{K}%
\!+\!\frac{\widehat{\Theta }}{3}\rangle \!\left[ \frac{2}{3}D_{k}\langle 
\widehat{\Theta }\rangle \!-\!D^{b}\langle \widehat{\sigma }_{kb}\rangle
\!+\!\frac{2\Theta }{3}\langle \widehat{K}_{k}\rangle \right] \!\!-\!\langle 
\widehat{\sigma }_{b}^{a}\rangle D^{b}\langle \widehat{\sigma }_{ka}\rangle
\!\! \\
+\frac{\langle \widehat{K}_{k}\rangle }{3}\!D^{a}\!\langle \widehat{K}%
_{a}\rangle \!\!+\!\varepsilon _{k}^{\ \ ab}\langle \widehat{K}_{c}\rangle
\!\langle \widehat{\sigma }_{b}^{c}\rangle \omega _{a}\!\!\!-\!\frac{\langle 
\widehat{\sigma }_{k}^{a}\rangle }{3}\!D_{a}\!\langle \widehat{\Theta }%
\rangle \!\!+\!\frac{\widetilde{\kappa }^{4}q^{a}}{4}\!\!\bigl[%
D_{a}q_{k}\!\!-\!2D_{k}q_{a}\!\!-\!\dot{\pi}_{\langle ka\rangle
}\!\!-\!\sigma _{ck}\pi _{a}^{c}\!-\!2A_{\langle k}q_{a\rangle }\! \\
+\sigma _{ba}\pi _{k}^{b}\!-\!\varepsilon _{cab}\omega ^{b}\pi _{k}^{c}\bigr]%
\!-\frac{\widetilde{\kappa }^{4}\!\pi ^{ab}}{4}\!\biggl[\!\frac{2q_{k}}{3}%
\sigma _{ab}\!-\!D_{k}\pi _{ab}\!\biggr]\!+\frac{\widetilde{\kappa }%
^{4}\left( \rho \!+\!p\right) }{6}\!\!\biggl[D_{k}\rho \!-\!\frac{3}{2}%
D^{b}\pi _{kb}\!-\!\Theta q_{k}\!\biggr]\!\!+\frac{\widetilde{\kappa }^{4}}{%
12}\! \\
\times \biggl[\!q_{k}D^{a}q_{a}\!-\!3\pi _{b}^{a}D^{b}\pi _{ka}\!-\!\pi
_{k}^{a}D_{a}\rho \!-\!3\varepsilon _{k}^{\ \ ab}q_{c}\omega _{a}\pi _{b}^{c}%
\biggr]\!+\!\frac{2\widetilde{\kappa }^{2}}{3}\!\biggl[\frac{D_{k}\langle 
\widetilde{\rho }\!+\!3\widetilde{\pi }\!-\!3\widetilde{p}\rangle }{4}%
+\!\langle \widehat{K}\!-\!\frac{\widehat{\Theta }}{3}\rangle \langle 
\widetilde{\pi }_{k}\rangle \! \\
-\langle \widetilde{\pi }_{\langle k\rangle }^{\prime }\rangle \!-\!\langle
K\rangle \langle \widetilde{q}_{k}\rangle \!-\!\langle \widetilde{q}\rangle
\langle 2\widehat{K}_{k}\!+\!K_{k}\rangle +\!\langle \widetilde{\pi }%
_{ka}\rangle \langle \widehat{A}^{a}\rangle \!-\!\langle \widetilde{\pi }%
\!-\!\widetilde{p}\rangle \langle \widehat{A}_{k}\rangle -\!\frac{5\langle 
\widehat{\sigma }_{ka}\rangle }{2}\langle \widetilde{\pi }^{a}\rangle -\!%
\frac{\widetilde{\kappa }^{2}q_{k}}{2}\Delta \widetilde{q}\! \\
+\frac{\widetilde{\kappa }^{2}\!\left( 2\lambda \!\!-\!\rho \!-\!3p\right) }{%
12}\Delta \widetilde{\pi }_{k}\!+\!\frac{5\widetilde{\kappa }^{2}\pi _{ka}}{8%
}\Delta \widetilde{\pi }^{a}\!\!-\!\frac{\Delta K}{4}~\!\Delta \widetilde{q}%
_{k}\!-\!\frac{\Delta K_{k}}{4}\Delta \widetilde{q}\!+\!\frac{\Delta 
\widehat{A}^{a}}{4}\Delta \widetilde{\pi }_{ka}\!\!-\!\frac{\!\left[ \Delta
\left( \widetilde{\pi }\!-\!\widetilde{p}\right) \right] }{4}\Delta \widehat{%
A}_{k}\!\biggr]\!\!\ ,
\end{gather*}%
\begin{gather*}
0=\langle \dot{\mathcal{E}}\rangle \!-\!D^{a}\langle \widehat{\mathcal{E}}%
_{a}\rangle \!+\!\frac{4}{3}\Theta \langle \mathcal{E}\rangle \!+\!\langle 
\widehat{\mathcal{E}}_{ab}\rangle \sigma ^{ab}\!-\!2\langle \widehat{%
\mathcal{E}}_{a}\rangle A^{a}\!+\!\langle \widehat{\sigma }^{ab}\rangle \!%
\biggl[\langle \dot{\widehat{\sigma }}_{\langle ab\rangle }\rangle
\!-\!D_{a}\langle \widehat{K}_{b}\rangle \!+\!\langle \widehat{K}\!+\!\frac{%
\widehat{\Theta }}{3}\rangle \sigma _{ab}\! \\
-2A_{b}\langle \widehat{K}_{a}\rangle \!+\!\frac{\Theta }{3}\langle \widehat{%
\sigma }_{ab}\rangle \!+\!\sigma _{ca}\langle \widehat{\sigma }%
_{b}^{c}\rangle \!\biggr]+\!\langle \widehat{K}^{a}\rangle \left[ \frac{2}{3}%
D_{a}\langle \widehat{\Theta }\rangle \!-\!D^{b}\langle \widehat{\sigma }%
_{ab}\rangle \!+\!\frac{2\Theta }{3}\langle \widehat{K}_{a}\rangle
\!-\!\langle \widehat{K}^{b}\rangle \sigma _{ab}\right] +\!\frac{\widetilde{%
\kappa }^{4}\pi ^{ab}}{4} \\
\times \!\left[ \dot{\pi}_{\langle ab\rangle
}\!+\!D_{a}q_{b}\!+\!2A_{b}q_{a}\!+\!\frac{\Theta }{3}\pi _{ab}\!+\!\left(
\rho \!+\!p\right) \sigma _{ab}\!+\!\sigma _{ca}\pi _{b}^{c}\right] +\!\frac{%
\widetilde{\kappa }^{4}q^{a}}{4}\!\!\biggl[D^{b}\pi _{ab}\!-\!\frac{2}{3}%
D_{a}\rho \!+\!\frac{2\Theta }{3}q_{a}\!-\!\sigma _{ab}q^{b}\!\biggr] \\
+\frac{\widetilde{\kappa }^{4}\left( \lambda \!+\!\rho \right) }{6}\Delta 
\widetilde{q}-\!\frac{2\widetilde{\kappa }^{2}}{3}\!\biggl[\frac{3\langle 
\widetilde{\rho }\!-\!\widetilde{\pi }\!+\!\widetilde{p}\rangle ^{\cdot }}{4}%
\!+\!D^{a}\langle \widetilde{q}_{a}\rangle \!+\!\Theta \langle \widetilde{%
\rho }\!+\!\widetilde{p}\rangle \!+\!\langle \widehat{\Theta }\rangle
\langle \widetilde{q}\rangle \!+\!2\langle \widetilde{q}_{a}\rangle
A^{a}\!+\!\langle \widetilde{\pi }_{ab}\rangle \sigma ^{ab}\!\biggr]\ .
\end{gather*}%
\[
0=\dot{\omega}_{\langle a\rangle }\!-\!\frac{1}{2}\varepsilon _{a}^{\ \
cd}D_{c}A_{d}\!+\!\frac{2\Theta }{3}\omega _{a}\!-\!\sigma _{ab}\omega ^{b}\
,
\]%
\begin{align*}
0& =\dot{\Theta}\!-\!D^{a}A_{a}\!+\!\frac{\Theta ^{2}}{3}\!-\!A^{a}A_{a}\!-%
\!2\omega _{a}\omega ^{a}\!+\!\sigma _{ab}\sigma ^{ab}\!+\!\frac{\kappa ^{2}%
}{2}\left( \rho \!+\!3p\right) \!-\!\Lambda \!+\!\frac{\widetilde{\kappa }%
^{4}\rho }{12}\left( 2\rho \!+\!3p\right)  \\
& \!-\!\frac{\widetilde{\kappa }^{4}}{4}q^{a}q_{a}\!+\!\langle \widehat{%
\Theta }\rangle \langle \widehat{K}\rangle \!-\!\langle \widehat{K}%
^{a}\rangle \langle \widehat{K}_{a}\rangle \!-\!\langle \mathcal{E}\rangle
\!+\!\frac{\widetilde{\kappa }^{2}}{2}\langle \widetilde{\rho }\!+\!%
\widetilde{\pi }\!+\!\widetilde{p}\rangle \ ,
\end{align*}%
\begin{gather*}
0=\dot{\sigma}_{\langle ab\rangle }\!-\!D_{\langle a}A_{b\rangle }\!+\!\frac{%
2\Theta }{3}\sigma _{ab}\!-\!A_{\langle a}A_{b\rangle }\!+\!\omega _{\langle
a}\omega _{b\rangle }\!+\!\sigma _{c\langle a}\sigma _{b\rangle
}^{c}\!+\!E_{ab}+\!\frac{1}{2}\langle \widehat{K}\!-\!\frac{\widehat{\Theta }%
}{3}\rangle \langle \widehat{\sigma }_{ab}\rangle +\!\frac{\langle \widehat{%
\mathcal{E}}_{ab}\rangle }{2} \\
-\frac{1}{2}\langle \widehat{K}_{\langle a}\rangle \langle \widehat{K}%
_{b\rangle }\rangle +\!\frac{1}{2}\langle \widehat{\sigma }_{c\langle
a}\rangle \langle \widehat{\sigma }_{b\rangle }^{c}\rangle \!\!-\!\frac{%
\widetilde{\kappa }^{2}}{3}\langle \widetilde{\pi }_{ab}\rangle +\!\frac{%
\widetilde{\kappa }^{4}}{8}\!\left[ \pi _{c\langle a}\pi _{b\rangle
}^{c}\!-\!q_{\langle a}q_{b\rangle }\!-\!\frac{\left( 2\lambda \!-\!\rho
\!-\!3p\right) }{3}\pi _{ab}\right] \!\!\!\ ,
\end{gather*}%
The constraint equations arising as averages are:%
\begin{gather*}
0\!=D^{a}H_{ak}\!\!-\!\frac{4\langle \mathcal{E}\rangle }{3}\!\omega
_{k}\!+\!3E_{ka}\omega ^{a}\!\!+\frac{\omega ^{a}}{2}\langle \widehat{%
\mathcal{E}}_{ak}\!\!\rangle +\!\frac{\varepsilon _{k}^{\ \ ab}}{2}\!\biggl[%
D_{a}\!\langle \widehat{\mathcal{E}}_{b}\rangle -2E_{ac}\sigma
_{b}^{c}+\langle \widehat{\mathcal{E}}_{ac}\rangle \sigma _{b}^{c}-\!\langle 
\widehat{\sigma }_{a}^{c}\!\rangle D_{b}\langle \widehat{K}_{c}\!\rangle  \\
\!-\frac{2\langle \widehat{K}_{b}\rangle }{3}\!D_{a}\!\langle \widehat{%
\Theta }\!\rangle -\!\frac{2\langle \widehat{\Theta }\rangle }{3}%
\!D_{a}\!\langle \widehat{K}_{b}\!\rangle +\sigma _{ac}\langle \widehat{K}%
^{c}\!\rangle \langle \widehat{K}_{b}\!\rangle +\!\langle \widehat{\sigma }%
_{da}\!\rangle \langle \widehat{\sigma }_{c}^{d}\rangle \sigma
_{b}^{c}\!-\!\langle \widehat{K}^{c}\rangle \!D_{b}\langle \widehat{\sigma }%
_{ac}\rangle \biggr]\!-\!\frac{\omega _{c}}{2}\!\langle \widehat{K}%
_{k}\!\rangle \!\langle \widehat{K}^{c}\rangle \! \\
+\frac{1}{2}\!\langle \widehat{K}-\frac{\widehat{\Theta }}{3}\rangle \!\left[
\langle \widehat{\sigma }_{k}^{c}\rangle \omega _{c}\!+\!\varepsilon _{k}^{\
\ ab}\!\langle \widehat{\sigma }_{ac}\rangle \sigma _{b}^{c}\right] \!+\!%
\frac{2\langle \widehat{\Theta }\rangle }{3}\langle \widehat{K}+\frac{%
\widehat{\Theta }}{3}\rangle \omega _{k}\!-\!\frac{\omega _{k}}{2}\!\!\left[
\langle \widehat{K}_{a}\rangle \!\langle \widehat{K}^{a}\rangle \!+\!\langle 
\widehat{\sigma }_{ab}\rangle \!\langle \widehat{\sigma }^{ab}\rangle \right]
\! \\
+\frac{\omega ^{a}}{2}\!\langle \widehat{\sigma }_{ca}\rangle \!\langle 
\widehat{\sigma }_{k}^{c}\rangle \!\!+\frac{\widetilde{\kappa }%
^{4}\varepsilon _{k}^{\ \ ab}}{8}\!\biggl[\frac{2}{3}q_{b}D_{a}\rho \!+\!%
\frac{2\!\left( \lambda +\rho \right) }{3}\!D_{a}q_{b}\!+\!q^{c}\!D_{b}\pi
_{ac}+\!\sigma _{ac}q^{c}q_{b}\!+\!\pi _{da}\pi _{c}^{d}\sigma _{b}^{c}+\pi
_{a}^{c}\!D_{b}q_{c}\!\biggr] \\
\!+\!\frac{\widetilde{\kappa }^{4}}{8}\biggl[\frac{4\!\left( \lambda +\rho
\right) }{3}\!\left( \rho +p\right) \!\omega _{k}\!-\!\frac{\left( 2\lambda
-\rho -3p\right) }{3}\!\left( \pi _{k}^{c}\omega _{c}\!+\!\varepsilon
_{k}^{\ \ ab}\!\pi _{ac}\sigma _{b}^{c}\right) \!-\!q_{a}q^{a}\omega
_{k}\!-\!q^{c}\omega _{c}q_{k}\! \\
\!-\pi _{ab}\pi ^{ab}\omega _{k}+\!\pi _{ca}\pi _{k}^{c}\omega ^{a}\biggr]%
\!-\!\frac{\widetilde{\kappa }^{2}}{3}\langle \widetilde{\pi }_{ka}\rangle
\omega ^{a}\!+\!\frac{\widetilde{\kappa }^{2}}{3}\varepsilon _{k}^{\ \
ab}\!D_{a}\langle \widetilde{q}_{b}\rangle \!+\!\frac{2\widetilde{\kappa }%
^{2}}{3}\!\langle \widetilde{\rho }+\widetilde{p}\rangle \omega _{k}\!-\!%
\frac{\widetilde{\kappa }^{2}}{3}\!\varepsilon _{k}^{\ \ ab}\!\langle 
\widetilde{\pi }_{a}^{c}\rangle \sigma _{bc}\ ,
\end{gather*}%
\begin{eqnarray*}
0 &=&D^{b}\sigma _{ab}-\frac{2D_{a}\Theta \!}{3}+\!\varepsilon _{a}^{\ \
ck}D_{c}\omega _{k}+2\varepsilon _{a}^{\ \ ck}A_{c}\omega _{k}\!+\!\frac{%
\widetilde{\kappa }^{4}\left( \lambda \!+\!\rho \right) }{6}q_{a}\! \\
&&-\frac{\widetilde{\kappa }^{4}q^{b}}{4}\pi _{ab}-\frac{2\langle \widehat{%
\Theta }\rangle }{3}\langle \widehat{K}_{a}\rangle +\langle \widehat{\sigma }%
_{ab}\rangle \langle \widehat{K}^{b}\rangle +\langle \widehat{\mathcal{E}}%
_{a}\rangle +\frac{2\widetilde{\kappa }^{2}}{3}\langle \widetilde{q}%
_{a}\rangle \ ,
\end{eqnarray*}%
\[
0=D^{a}\omega _{a}\!-\!A_{a}\omega ^{a}\ ,
\]%
\[
0=D_{\langle c}\omega _{k\rangle }\!+\!\varepsilon _{ab\langle k}D^{b}\sigma
_{c\rangle }^{a}\!+\!2A_{\langle c}\omega _{k\rangle }\!+\!H_{ab}\ ,
\]%
\begin{gather*}
0=D^{a}E_{ak}\!-\!\frac{1}{2}D^{a}\langle \widehat{\mathcal{E}}_{ak}\rangle
\!+\!\frac{1}{3}D_{k}\langle \mathcal{E}\rangle \!-\!3H_{ka}\omega
^{a}\!+\!\varepsilon _{k}^{\ \ ab}H_{ac}\sigma _{b}^{c}\!+\!\frac{\Theta }{3}%
\langle \widehat{\mathcal{E}}_{k}\rangle \!-\!\frac{1}{2}\left( 3\omega
_{ka}\!+\!\sigma _{ka}\right) \!\langle \widehat{\mathcal{E}}^{a}\rangle \!
\\
\!-\frac{2\langle \widehat{\Theta }\rangle }{9}D_{k}\langle \widehat{\Theta }%
\rangle \!-\!\frac{\langle \widehat{\sigma }_{ak}\rangle }{2}D^{a}\langle 
\widehat{K}\!-\!\frac{\widehat{\Theta }}{3}\rangle \!-\!\frac{1}{2}\langle 
\widehat{K}\!-\!\frac{\widehat{\Theta }}{3}\rangle D^{a}\langle \widehat{%
\sigma }_{ak}\rangle \!-\!\frac{\langle \widehat{\sigma }_{a}^{c}\rangle }{2}%
D^{a}\langle \widehat{\sigma }_{ck}\rangle \!+\!\frac{\langle \widehat{%
\Theta }\rangle }{3}\langle \widehat{K}^{b}\rangle \sigma _{kb}\! \\
\!-\frac{\langle \widehat{\sigma }_{a}^{b}\rangle }{2}\langle \widehat{K}%
^{a}\rangle \sigma _{kb}\!+\!\frac{2\langle \widehat{\sigma }^{ab}\rangle }{3%
}D_{k}\langle \widehat{\sigma }_{ab}\rangle \!-\!\frac{\langle \widehat{K}%
^{a}\rangle }{3}D_{k}\langle \widehat{K}_{a}\rangle +\!\frac{\langle 
\widehat{K}_{a}\rangle }{2}D^{a}\langle \widehat{K}_{k}\rangle \!+\!\frac{%
\langle \widehat{K}_{k}\rangle }{2}D^{a}\langle \widehat{K}_{a}\rangle \! \\
+\frac{3}{2}\varepsilon _{k}^{\ \ ad}\langle \widehat{K}_{c}\rangle \langle 
\widehat{\sigma }_{d}^{c}\rangle \omega _{a}\!+\!\langle \widehat{\Theta }%
\rangle \varepsilon _{k}^{\ \ cd}\langle \widehat{K}_{c}\rangle \omega
_{d}\!+\!\frac{\Theta }{3}\langle \widehat{\sigma }_{k}^{a}\rangle \langle 
\widehat{K}_{a}\rangle \!-\!\frac{2\Theta }{9}\langle \widehat{\Theta }%
\rangle \langle \widehat{K}_{k}\rangle \!-\!\langle \widehat{\sigma }%
_{kb}\rangle D^{a}\langle \widehat{\sigma }_{a}^{b}\rangle \!-\!\frac{%
\widetilde{\kappa }^{4}}{8} \\
\times \!\biggl[\frac{4\left( \lambda \!+\!\rho \right) }{9}D_{k}\rho \!+\!%
\frac{\pi _{ak}}{3}D^{a}\left( \rho \!+\!3p\right) \!-\!\frac{\left(
2\lambda \!-\!\rho \!-\!3p\right) }{3}D^{a}\pi _{ak}\!+\!\pi
_{a}^{c}D^{a}\pi _{ck}\!\!-\!\frac{4\pi ^{ab}}{3}D_{k}\pi _{ab}\!+\!\pi
_{kb}D^{a}\pi _{a}^{b}\! \\
-q_{a}D^{a}q_{k}\!-\!q_{k}D^{a}q_{a}\!+\!3\varepsilon _{k}^{\ \
ad}q_{c}\omega _{a}\pi _{d}^{c}\!-\!\pi _{a}^{b}q^{a}\sigma _{kb}\!+\frac{%
2\left( \lambda \!+\!\rho \right) }{3}\!\!\left( \sigma _{kb}q^{b}\!-\!\frac{%
2\Theta }{3}q_{k}\!+\!3\varepsilon _{k}^{\ \ cd}q_{c}\omega _{d}\right)  \\
\!+\frac{2q^{a}}{3}D_{k}q_{a}+\!\frac{2\Theta }{3}\pi _{k}^{a}q_{a}\!\biggr]%
\!\!+\!\frac{\widetilde{\kappa }^{2}}{3}D^{a}\langle \widetilde{\pi }%
_{ak}\rangle \!\!-\!\frac{\widetilde{\kappa }^{2}}{6}D_{k}\langle \widetilde{%
\rho }\!-\!\widetilde{\pi }\!+\!\widetilde{p}\rangle \!+\!\frac{2\widetilde{%
\kappa }^{2}}{9}\Theta \langle \widetilde{q}_{k}\rangle \!-\!\frac{%
\widetilde{\kappa }^{2}}{3}\langle \widetilde{q}^{a}\rangle \left( 3\omega
_{ka}\!+\!\sigma _{ka}\right) \ .
\end{gather*}

\section{The difference equations}

The evolution equations arising as differences are:%
\[
\dot{\rho}\!+\!\left( \rho \!+\!p\right) \Theta
\!+\!D^{a}q_{a}\!+\!2q^{a}A_{a}\!+\!\pi _{ab}\sigma ^{ab}=\!-\!\Delta 
\widetilde{q}\ , 
\]%
\[
\dot{q}_{\langle a\rangle }\!+\!D_{a}p\!+\!D^{b}\pi _{ab}\!+\!\frac{4}{3}%
\Theta q_{a}\!+\!\sigma _{ab}q^{b}\!-\!\omega _{ab}q^{b}\!+\!\left( \rho
\!+\!p\right) A_{a}\!+\!\pi _{ab}A^{b}=\!-\!\Delta \widetilde{\pi }_{a}\ , 
\]%
\begin{gather*}
0=\Delta \dot{\mathcal{E}}\!-\!D^{a}\Delta \widehat{\mathcal{E}}_{a}\!+\!%
\frac{4}{3}\Theta \Delta \mathcal{E}+\sigma ^{ab}\!\Delta \widehat{\mathcal{E%
}}_{ab}-\!2A^{a}\!\Delta \widehat{\mathcal{E}}_{a}+\!\widetilde{\kappa }%
^{2}\!\biggl[\pi ^{ab}D_{a}\langle \widehat{K}_{b}\rangle \!-\!\langle 
\widehat{\sigma }^{ab}\rangle D_{a}q_{b}\!-\!q^{a}D^{b}\langle \widehat{%
\sigma }_{ab}\rangle \! \\
+\langle \widehat{K}^{a}\rangle D^{b}\pi _{ab}\!+\!\frac{2q^{a}}{3}%
D_{a}\langle \widehat{\Theta }\rangle \!-\!\frac{2\langle \widehat{K}%
^{a}\rangle }{3}D_{a}\rho \!-\!\pi ^{ab}\langle \dot{\widehat{\sigma }}%
_{\langle ab\rangle }\rangle \!\!-\!\dot{\pi}_{\langle ab\rangle }\langle 
\widehat{\sigma }^{ab}\rangle \!-\!\!2A_{a}q_{b}\langle \widehat{\sigma }%
^{ab}\rangle \!+\!2A_{a}\pi ^{ab}\langle \widehat{K}_{b}\rangle \! \\
-\frac{2\Theta }{3}\pi _{ab}\langle \widehat{\sigma }^{ab}\rangle
\!-\!\left( \rho \!+\!p\right) \sigma _{ab}\langle \widehat{\sigma }%
^{ab}\rangle \!-\!\langle \widehat{K}\!+\!\frac{\widehat{\Theta }}{3}\rangle
\sigma _{ab}\pi ^{ab}\!-\!2\sigma _{ca}\pi ^{ab}\langle \widehat{\sigma }%
_{b}^{c}\rangle \!+\!\frac{4\Theta }{3}q^{a}\langle \widehat{K}_{a}\rangle
\!-\!2\sigma _{ab}q^{b}\langle \widehat{K}^{a}\rangle \! \\
-\frac{\left[ \Delta \!\left( \widetilde{\rho }\!-\!\widetilde{\pi }\!+\!%
\widetilde{p}\right) \right] ^{\cdot }}{2}\!\!-\!\frac{2}{3}\!D^{a}\!\Delta 
\widetilde{q}_{a}\!\!-\!\frac{2\left[ \Delta \!\left( \widetilde{\rho }\!+\!%
\widetilde{p}\right) \right] \!}{3}\Theta \!+\!\frac{2\widetilde{\kappa }%
^{2}\!\left( \lambda \!+\!\rho \right) }{3}\langle \widetilde{q}\rangle
\!\!-\!\frac{2\langle \widehat{\Theta }\rangle }{3}\Delta \widetilde{q}\!-\!%
\frac{4A^{a}}{3}\Delta \widetilde{q}_{a}\!\!-\!\frac{2\sigma ^{ab}}{3}\Delta 
\widetilde{\pi }_{ab}\!\biggr],
\end{gather*}%
\begin{gather*}
0=\Delta \dot{\widehat{\mathcal{E}}}_{\langle k\rangle }\!+\!\frac{4}{3}%
\Theta \Delta \widehat{\mathcal{E}}_{k}-\!\frac{1}{3}D_{k}\Delta \mathcal{E}%
\!-\!\frac{4}{3}A_{k}\!\Delta \mathcal{E}-\!D^{a}\Delta \widehat{\mathcal{E}}%
_{ka}\!-\!\Delta \widehat{\mathcal{E}}_{ka}A^{a}\!-\!\left( \omega
_{ka}\!-\!\sigma _{ka}\right) \Delta \widehat{\mathcal{E}}^{a}\!+\!%
\widetilde{\kappa }^{2}\! \\
\times \biggl[\left( \rho \!+\!p\right) D^{b}\langle \widehat{\sigma }%
_{kb}\rangle \!+\!\langle \widehat{K}\!+\!\frac{\widehat{\Theta }}{3}\rangle
D^{b}\pi _{kb}\!-\!\frac{2\left( \rho \!+\!p\right) }{3}D_{k}\langle 
\widehat{\Theta }\rangle \!\!-\!\frac{2}{3}\langle \widehat{K}\!+\!\frac{%
\widehat{\Theta }}{3}\rangle D_{k}\rho \!-\!2q^{a}D_{k}\langle \widehat{K}%
_{a}\rangle \\
\!-2\langle \widehat{K}^{a}\rangle D_{k}q_{a}\!+\!q^{a}D_{a}\langle \widehat{%
K}_{k}\rangle \!+\!\langle \widehat{K}^{a}\rangle D_{a}q_{k}\!\!+\!\frac{%
\langle \widehat{K}_{k}\rangle }{3}D^{a}q_{a}-\!\pi ^{ab}D_{k}\langle 
\widehat{\sigma }_{ab}\rangle \!-\!\langle \widehat{\sigma }^{ab}\rangle
D_{k}\pi _{ab} \\
\!+\pi _{b}^{a}D^{b}\langle \widehat{\sigma }_{ka}\rangle \!+\!\langle 
\widehat{\sigma }_{b}^{a}\rangle D^{b}\pi _{ka}\!+\!\frac{\pi _{k}^{a}}{3}%
D_{a}\langle \widehat{\Theta }\rangle \!+\!\frac{\langle \widehat{\sigma }%
_{k}^{a}\rangle }{3}D_{a}\rho \!-\!\dot{\pi}_{\langle ka\rangle }\langle 
\widehat{K}^{a}\rangle \!+\!q^{a}\langle \dot{\widehat{\sigma }}_{\langle
ka\rangle }\rangle \!+\!q^{a}\sigma _{ck}\langle \widehat{\sigma }%
_{a}^{c}\rangle \\
\!-\sigma _{ck}\pi _{a}^{c}\langle \widehat{K}^{a}\rangle \!\!-\!\frac{%
2\left( \rho \!+\!p\right) }{3}\Theta \langle \widehat{K}_{k}\rangle \!+\!%
\frac{2\Theta }{3}\langle \widehat{K}\!+\!\frac{\widehat{\Theta }}{3}\rangle
q_{k}\!-\!2q^{a}A_{\langle k}\langle \widehat{K}_{a\rangle }\rangle
\!-\!2A_{\langle k}q_{a\rangle }\langle \widehat{K}^{a}\rangle \!+\!\sigma
_{ba}\pi _{k}^{b}\langle \widehat{K}^{a}\rangle \! \\
-\sigma _{ba}q^{a}\langle \widehat{\sigma }_{k}^{b}\rangle \!+\!\frac{2}{3}%
q_{k}\sigma _{ab}\langle \widehat{\sigma }^{ab}\rangle \!-\!\frac{2}{3}%
\sigma _{ab}\pi ^{ab}\langle \widehat{K}_{k}\rangle \!+\!\varepsilon
_{cab}q^{a}\omega ^{b}\langle \widehat{\sigma }_{k}^{c}\rangle
\!-\!\varepsilon _{cab}\omega ^{b}\pi _{k}^{c}\langle \widehat{K}^{a}\rangle
\!+\!\varepsilon _{k}^{\ \ ab}q_{c}\omega _{a}\langle \widehat{\sigma }%
_{b}^{c}\rangle \! \\
-\varepsilon _{k}^{\ \ ab}\omega _{a}\pi _{b}^{c}\langle \widehat{K}%
_{c}\rangle \!\!-\!\frac{2}{3}\Delta \widetilde{\pi }_{\langle k\rangle
}^{\prime }\!+\!\frac{1}{6}D_{k}\!\Delta \left( \widetilde{\rho }\!+\!3%
\widetilde{\pi }\!-\!3\widetilde{p}\right) \!+\!\frac{2\widetilde{\kappa }%
^{2}}{9}\left( 2\lambda \!-\!\rho \!-\!3p\right) \!\langle \widetilde{\pi }%
_{k}\rangle \!+\!\frac{2}{3}\langle \widehat{K}\!-\!\frac{\widehat{\Theta }}{%
3}\rangle \Delta \widetilde{\pi }_{k}\! \\
+\frac{q_{k}}{3}D^{a}\langle \widehat{K}_{a}\rangle \!\!-\!\frac{2\langle 
\widetilde{q}_{k}\rangle }{3}\!\Delta K\!-\!\frac{2\langle K\rangle }{3}%
\Delta \widetilde{q}_{k}\!-\!\frac{2}{3}\langle 2\widehat{K}%
_{k}\!+\!K_{k}\rangle \Delta \widetilde{q}\!-\!\frac{2\langle \widetilde{q}%
\rangle }{3}\!\left( 2\widetilde{\kappa }^{2}q_{k}\!+\!\Delta K_{k}\right)
\!+\!\frac{2\langle \widehat{A}^{a}\rangle }{3}\!\Delta \widetilde{\pi }_{ka}
\\
+\frac{2\langle \widetilde{\pi }_{ka}\rangle }{3}\Delta \widehat{A}^{a}-\!%
\frac{2\langle \widehat{A}_{k}\rangle }{3}\!\Delta \left( \widetilde{\pi }%
\!-\!\widetilde{p}\right) \!-\!\frac{2\langle \widetilde{\pi }\!-\!%
\widetilde{p}\rangle }{3}\Delta \widehat{A}_{k}\!+\!\frac{5\widetilde{\kappa 
}^{2}}{3}\pi _{ka}\langle \widetilde{\pi }^{a}\rangle \!-\!\frac{5\langle 
\widehat{\sigma }_{ka}\rangle }{3}\Delta \widetilde{\pi }^{a}\!\biggr]\ ,
\end{gather*}%
\begin{gather*}
0=\!\Delta \dot{\widehat{\mathcal{E}}}_{\langle kj\rangle }\!\!-\!D_{\langle
k}\Delta \widehat{\mathcal{E}}_{j\rangle }\!+\!\frac{\Theta }{3}\Delta 
\widehat{\mathcal{E}}_{kj}\!-\!2A_{\langle k}\!\Delta \widehat{\mathcal{E}}%
_{j\rangle }+\!\frac{4\sigma _{kj}\!}{3}\Delta \mathcal{E\!}-\!\left( \omega
_{a\langle k}\!-\!\sigma _{a\langle k}\right) \!\Delta \widehat{\mathcal{E}}%
_{j\rangle }^{a}\!+\!\widetilde{\kappa }^{2}\!\biggl\{\frac{2\langle 
\widehat{\Theta }\rangle }{3}D_{\langle k}q_{j\rangle } \\
\!\!-\langle \widehat{K}\!-\!\frac{\widehat{\Theta }}{3}\rangle ^{\cdot }\pi
_{kj}\!\!-\!2\pi _{\langle j}^{a}\langle \dot{\widehat{\sigma }}_{k\rangle
a}\rangle \!\!-\!2\langle \widehat{\sigma }_{\langle j}^{a}\rangle \dot{\pi}%
_{k\rangle a}\!\!-\!\frac{\left( \rho \!+\!3p\right) ^{\cdot }}{3}\langle 
\widehat{\sigma }_{kj}\rangle \!\!-\!2q_{\langle k}D_{j\rangle }\!\langle 
\widehat{K}\!-\!\widehat{\Theta }\rangle \!\!-\!\frac{2\langle \widehat{K}%
_{\langle k}\rangle }{3}D_{j\rangle }\!\!\left( \rho \!-\!3p\right) \\
\!\!+\pi _{\langle j}^{a}D_{k\rangle }\langle \widehat{K}_{a}\rangle
\!\!-\!\langle \widehat{\sigma }_{\langle j}^{a}\rangle D_{k\rangle
}q_{a}\!-\!2q_{\langle k}D^{b}\langle \widehat{\sigma }_{j\rangle b}\rangle
\!+\!2\langle \widehat{K}_{\langle k}\rangle D^{b}\pi _{j\rangle
b}+\!\langle \widehat{K}_{a}\rangle D_{\langle k}\pi _{j\rangle }^{a}\!+\!%
\frac{14\Theta }{3}q_{\langle k}\langle \widehat{K}_{j\rangle }\rangle \\
\!\!-\langle \widehat{K}\!-\!\frac{\widehat{\Theta }}{3}\rangle \!\left[ 
\dot{\pi}_{\langle kj\rangle }\!+\!\frac{\Theta }{3}\pi _{kj}\!+\!\pi
_{\langle k}^{a}\left( \omega _{j\rangle a}\!+\!\sigma _{j\rangle a}\right)
\!+\!2q_{\langle k}A_{j\rangle }\right] \!\!-\!\pi _{\langle j}^{a}\omega
_{k\rangle c}\langle \widehat{\sigma }_{a}^{c}\rangle +\!q_{\langle k}\sigma
_{j\rangle b}\langle \widehat{K}^{b}\rangle \\
\!+\frac{\left( 2\lambda \!-\!\rho \!-\!3p\right) }{3}\!\left[ \langle \dot{%
\widehat{\sigma }}_{\langle kj\rangle }\rangle \!+\!\frac{\Theta }{3}\langle 
\widehat{\sigma }_{kj}\rangle \!+\!\langle \widehat{\sigma }_{\langle
k}^{a}\rangle \!\left( \omega _{j\rangle a}\!+\!\sigma _{j\rangle a}\right)
\!\!-\!2\langle \widehat{K}_{\langle k}\rangle A_{j\rangle }\right] \!\!\!+\!%
\frac{2\left( \rho \!+\!p\right) }{3}\langle \widehat{\Theta }\rangle \sigma
_{kj} \\
\!+\langle \widehat{K}_{\langle k}\rangle \sigma _{j\rangle b}q^{b}\!\!-\!%
\frac{2\Theta }{3}\pi _{\langle j}^{a}\langle \widehat{\sigma }_{k\rangle
a}\rangle \!\!-\!\frac{2\left( \lambda \!+\!\rho \right) }{3}\!\!\biggl[%
D_{\langle k}\langle \widehat{K}_{j\rangle }\rangle \!-\!\langle \widehat{K}%
\!+\!\frac{\widehat{\Theta }}{3}\rangle \sigma _{kj}\biggr]%
\!\!\!-q_{a}D_{\langle k}\langle \widehat{\sigma }_{j\rangle }^{a}\rangle
\!\!-\!\langle \widehat{\sigma }_{\langle j}^{a}\rangle \omega _{k\rangle
c}\pi _{a}^{c}\! \\
\!\!\!+2\sigma _{jk}q^{a}\langle \widehat{K}_{a}\rangle \!\!-\!3q_{\langle
k}\omega _{j\rangle a}\langle \widehat{K}^{a}\rangle \!\!-\!3\langle 
\widehat{K}_{\langle k}\rangle \omega _{j\rangle a}q^{a}\!-\!2q_{\langle
k}\langle \widehat{\sigma }_{j\rangle a}\rangle A^{a}\!+\!2\langle \widehat{K%
}_{\langle k}\rangle \pi _{j\rangle a}A^{a}\!\!+\!2\pi _{a\langle
j}A_{k\rangle }\!\langle \widehat{K}^{a}\rangle \\
\!-2\langle \widehat{\sigma }_{a\langle j}\rangle A_{k\rangle }q^{a}\!-\!\pi
_{c}^{a}\langle \widehat{\sigma }_{\langle k}^{c}\rangle \sigma _{j\rangle
a}\!-\!\langle \widehat{\sigma }_{c}^{a}\rangle \pi _{\langle k}^{c}\sigma
_{j\rangle a}\!+\!\frac{2}{3}\Delta \dot{\widetilde{\pi }}_{\langle
kj\rangle }\!+\!\frac{8A_{\langle k}}{3}\Delta \widetilde{q}_{j\rangle }\!+\!%
\frac{2}{3}D_{\langle k}\Delta \widetilde{q}_{j\rangle } \\
\!-\frac{2\langle \widehat{K}_{\langle k}\rangle }{3}\Delta \widetilde{\pi }%
_{j\rangle }\!-\!\frac{2\widetilde{\kappa }^{2}}{3}\langle \widetilde{\pi }%
_{\langle k}\rangle q_{j\rangle }\!+\!\frac{2\!\left[ \Delta \left( 
\widetilde{\rho }\!+\!\widetilde{p}\right) \right] }{3}\sigma _{jk}\!+\!%
\frac{2\Theta }{9}\Delta \widetilde{\pi }_{jk}-\!\frac{2}{3}\left( \omega
_{a\langle k}\!-\!\sigma _{a\langle k}\right) \!\Delta \widetilde{\pi }%
_{j\rangle }^{a}\!\biggr\}\ .
\end{gather*}

The constraint equations arising as differences are:%
\begin{gather*}
0=D^{a}\Delta \widehat{\mathcal{E}}_{ak}\!-\!\frac{2}{3}D_{k}\Delta \mathcal{%
E}-\!\frac{2\Theta }{3}\Delta \widehat{\mathcal{E}}_{k}\!+\!\left( 3\omega
_{ka}\!+\!\sigma _{ka}\right) \Delta \widehat{\mathcal{E}}^{a}\!+\!%
\widetilde{\kappa }^{2}\!\biggl[\!-\!\frac{4\left( \lambda \!+\!\rho \right) 
}{9}D_{k}\langle \widehat{\Theta }\rangle \!-\!\frac{4}{9}\langle \widehat{%
\Theta }\rangle D_{k}\rho \\
\!-\pi _{ak}D^{a}\langle \widehat{K}\!-\!\frac{\widehat{\Theta }}{3}\rangle
\!-\!\frac{\langle \widehat{\sigma }_{ak}\rangle }{3}D^{a}\left( \rho
\!+\!3p\right) \!+\!\frac{\left( 2\lambda \!-\!\rho \!-\!3p\right) }{3}%
D^{a}\langle \widehat{\sigma }_{ak}\rangle \!-\!\langle \widehat{K}\!-\!%
\frac{\widehat{\Theta }}{3}\rangle D^{a}\pi _{ak}\!-\!\pi
_{a}^{c}D^{a}\langle \widehat{\sigma }_{ck}\rangle \! \\
\!-\langle \widehat{\sigma }_{a}^{c}\rangle D^{a}\pi _{ck}\!+\!\frac{4\pi
^{ab}}{3}D_{k}\langle \widehat{\sigma }_{ab}\rangle \!\!+\!\frac{4\langle 
\widehat{\sigma }^{ab}\rangle }{3}D_{k}\pi _{ab}\!-\!\pi _{kb}D^{a}\langle 
\widehat{\sigma }_{a}^{b}\rangle \!\!+\!3\varepsilon _{k}^{\ \ ad}\left[
\langle \widehat{K}_{c}\rangle \omega _{a}\pi _{d}^{c}-q_{c}\omega
_{a}\langle \widehat{\sigma }_{d}^{c}\rangle \right] \!\!\! \\
-q_{a}D^{a}\langle \widehat{K}_{k}\rangle \!-\pi _{a}^{b}\langle \widehat{K}%
^{a}\rangle \sigma _{kb}\!-\!\langle \widehat{K}_{a}\rangle
D^{a}q_{k}\!-\!q_{k}D^{a}\langle \widehat{K}_{a}\rangle \!-\!\langle 
\widehat{K}_{k}\rangle D^{a}q_{a}\!-\!\langle \widehat{\sigma }_{kb}\rangle
D^{a}\pi _{a}^{b}\! \\
+\frac{2q^{a}}{3}D_{k}\langle \widehat{K}_{a}\rangle \!+\!\frac{2\langle 
\widehat{K}^{a}\rangle }{3}D_{k}q_{a}\!+\langle \widehat{\sigma }%
_{a}^{b}\rangle q^{a}\sigma _{kb}\!+\!\langle \widehat{\Theta }\rangle \!\!%
\left[ \frac{4\Theta }{9}q_{k}\!-\!\frac{2}{3}\sigma
_{kb}q^{b}\!\!-\!2\varepsilon _{k}^{\ \ cd}q_{c}\omega _{d}\right] \!\! \\
-\frac{2\left( \lambda \!+\!\rho \right) }{3}\!\left[ \frac{2\Theta }{3}%
\langle \widehat{K}_{k}\rangle \!-\!\sigma _{kb}\langle \widehat{K}%
^{b}\rangle \!-\!3\varepsilon _{k}^{\ \ cd}\langle \widehat{K}_{c}\rangle
\omega _{d}\right] \!+\frac{2\Theta }{3}\pi _{k}^{a}\langle \widehat{K}%
_{a}\rangle \!-\!\frac{2\Theta }{3}\langle \widehat{\sigma }_{k}^{a}\rangle
q_{a} \\
\!\!-\frac{2}{3}D^{a}\Delta \widetilde{\pi }_{ak}\!+\!\frac{1}{3}%
D_{k}\!\Delta \left( \widetilde{\rho }\!-\!\widetilde{\pi }\!+\!\widetilde{p}%
\right) \!-\!\frac{4\Theta }{9}\Delta \widetilde{q}_{k}\!+\!\frac{2}{3}%
\left( 3\omega _{ka}\!+\!\sigma _{ka}\right) \Delta \widetilde{q}^{a}\!%
\biggr]\ ,
\end{gather*}%
\[
0=\frac{\widetilde{\kappa }^{2}}{3}\left( \lambda \!-\!2\rho \!-\!3p\right)
\langle \widehat{\Theta }\rangle \!-\!\widetilde{\kappa }^{2}\left( \lambda
\!+\!\rho \right) \langle \widehat{K}\rangle \!-\!2\widetilde{\kappa }%
^{2}q^{a}\langle \widehat{K}_{a}\rangle \!-\!\Delta \mathcal{E}\!-\!\frac{%
\Delta \widetilde{\Lambda }}{2}\!+\!\frac{\widetilde{\kappa }^{2}}{2}\Delta
\left( \widetilde{\rho }\!+\!\widetilde{\pi }\!+\!\widetilde{p}\right) \ , 
\]%
\[
0=\frac{2\widetilde{\kappa }^{2}}{3}\left( \lambda \!+\!\rho \right) \langle 
\widehat{K}_{a}\rangle \!-\!\frac{2\widetilde{\kappa }^{2}}{3}\langle 
\widehat{\Theta }\rangle q_{a}\!-\!\widetilde{\kappa }^{2}\pi _{ab}\langle 
\widehat{K}^{b}\rangle \!+\!\widetilde{\kappa }^{2}q^{b}\langle \widehat{%
\sigma }_{ab}\rangle \!+\!\Delta \widehat{\mathcal{E}}_{a}\!+\!\frac{2%
\widetilde{\kappa }^{2}}{3}\Delta \widetilde{q}_{a}\ , 
\]%
\[
0=\frac{\widetilde{\kappa }^{2}}{6}\left( 2\lambda \!-\!\rho \!-\!3p\right)
\langle \widehat{\sigma }_{ab}\rangle \!-\!\frac{\widetilde{\kappa }^{2}}{2}%
\langle \widehat{K}\!-\!\frac{\widehat{\Theta }}{3}\rangle \pi _{ab}\!-\!%
\widetilde{\kappa }^{2}q_{\langle a}\langle \widehat{K}_{b\rangle }\rangle
\!-\!\widetilde{\kappa }^{2}\pi _{c\langle a}\langle \widehat{\sigma }%
_{b\rangle }^{c}\rangle \!+\!\frac{\Delta \widehat{\mathcal{E}}_{ab}}{2}\!-\!%
\frac{\widetilde{\kappa }^{2}}{3}\Delta \widetilde{\pi }_{ab}\ , 
\]%
\begin{gather*}
0=\varepsilon _{k}^{\ \ ab}D_{a}\Delta \widehat{\mathcal{E}}_{b}-\!\frac{%
8\omega _{k}\!}{3}\Delta \mathcal{E}+\omega ^{a}\!\Delta \widehat{\mathcal{E}%
}_{ak}\!+\!\varepsilon _{k}^{\ \ ab}\sigma _{b}^{c}\!\Delta \widehat{%
\mathcal{E}}_{ac}+\!\widetilde{\kappa }^{2}\!\biggl[\frac{2}{3}\varepsilon
_{k}^{\ \ ab}\langle \widehat{K}_{b}\rangle D_{a}\rho \!-\!\frac{2}{3}%
\varepsilon _{k}^{\ \ ab}q_{b}D_{a}\langle \widehat{\Theta }\rangle \\
\!+\frac{2\left( \lambda \!+\!\rho \right) }{3}\varepsilon _{k}^{\ \
ab}D_{a}\langle \widehat{K}_{b}\rangle \!-\!\frac{2\langle \widehat{\Theta }%
\rangle }{3}\varepsilon _{k}^{\ \ ab}D_{a}q_{b}\!-\!\varepsilon
_{abk}q^{c}D^{b}\langle \widehat{\sigma }_{c}^{a}\rangle \!+\!\varepsilon
_{abk}\langle \widehat{K}^{c}\rangle D^{b}\pi _{c}^{a}\!+\!\varepsilon
_{k}^{\ \ \ ab}\pi _{a}^{c}D_{b}\langle \widehat{K}_{c}\rangle \\
\!-\varepsilon _{k}^{\ \ \ ab}\langle \widehat{\sigma }_{a}^{c}\rangle
D_{b}q_{c}\!+\!\frac{\left( 2\lambda \!-\!\rho \!-\!3p\right) }{3}\!\left[
\langle \widehat{\sigma }_{k}^{c}\rangle \omega _{c}\!+\!\varepsilon _{k}^{\
\ ab}\langle \widehat{\sigma }_{ac}\rangle \sigma _{b}^{c}\right]
\!-\!\langle \widehat{K}\!-\!\frac{\widehat{\Theta }}{3}\rangle \!\left[ \pi
_{k}^{c}\omega _{c}\!+\!\varepsilon _{k}^{\ \ ab}\pi _{ac}\sigma _{b}^{c}%
\right] \\
\!-\!\left[ \frac{4\left( \lambda \!+\!\rho \right) }{3}\langle \widehat{K}%
\!+\!\frac{\widehat{\Theta }}{3}\rangle \!+\!\frac{4\left( \rho
\!+\!p\right) }{3}\langle \widehat{\Theta }\rangle \!+\!2q_{a}\langle 
\widehat{K}^{a}\rangle \!-\!2\pi _{ab}\langle \widehat{\sigma }^{ab}\rangle %
\right] \!\omega _{k}\!-\!q^{c}\omega _{c}\langle \widehat{K}_{k}\rangle
\!-\!\langle \widehat{K}^{c}\rangle \omega _{c}q_{k} \\
\!+\varepsilon _{k}^{\ \ ac}\sigma _{ab}q^{b}\langle \widehat{K}_{c}\rangle
\!+\!\varepsilon _{k}^{\ \ ac}\sigma _{ab}\langle \widehat{K}^{b}\rangle
q_{c}\!-\!\pi _{ca}\langle \widehat{\sigma }_{k}^{c}\rangle \omega
^{a}\!-\!\langle \widehat{\sigma }_{ca}\rangle \pi _{k}^{c}\omega
^{a}\!-\!\varepsilon _{k}^{\ \ ab}\pi _{da}\langle \widehat{\sigma }%
_{c}^{d}\rangle \sigma _{b}^{c}\!-\!\varepsilon _{k}^{\ \ ab}\langle 
\widehat{\sigma }_{da}\rangle \pi _{c}^{d}\sigma _{b}^{c} \\
\!-\frac{2\omega ^{a}}{3}\Delta \widetilde{\pi }_{ka}\!+\!\frac{2}{3}%
\varepsilon _{k}^{\ \ ab}D_{a}\Delta \widetilde{q}_{b}\!+\!\frac{4\omega _{k}%
}{3}\!\Delta \left( \widetilde{\rho }\!+\!\widetilde{p}\right) -\!\frac{2}{3}%
\varepsilon _{k}^{\ \ ab}\sigma _{bc}\Delta \widetilde{\pi }_{a}^{c}\biggr]\
,
\end{gather*}%
\begin{gather*}
0=\varepsilon _{ab\langle k}D^{a}\Delta \widehat{\mathcal{E}}_{j\rangle
}^{b}\!\!-\!\varepsilon _{\langle k}^{\ \ \ ab}\sigma _{j\rangle a}\Delta 
\widehat{\mathcal{E}}_{b}\!-\!3\omega _{\langle k}\!\Delta \widehat{\mathcal{%
E}}_{j\rangle }\!+\!\!\widetilde{\kappa }^{2}\varepsilon _{\langle k}^{\ \ \
\ ab}\biggl[\pi _{j\rangle a}D_{b}\langle \widehat{K}\!-\!\frac{\widehat{%
\Theta }}{3}\rangle \!+\!\frac{\langle \widehat{\sigma }_{j\rangle a}\rangle 
}{3}D_{b}\left( \rho \!+\!3p\right) \\
-\sigma _{j\rangle b}\pi _{a}^{c}\langle \widehat{K}_{c}\rangle \!+\sigma
_{j\rangle b}\langle \widehat{\sigma }_{a}^{c}\rangle q_{c}-\!\frac{2\left(
\lambda \!+\!\rho \right) }{3}\sigma _{j\rangle a}\langle \widehat{K}%
_{b}\rangle \!+\!\frac{2\langle \widehat{\Theta }\rangle }{3}\sigma
_{j\rangle a}q_{b}-\!\frac{2\sigma _{j\rangle a}}{3}\Delta \widetilde{q}_{b}%
\biggr] \\
+\widetilde{\kappa }^{2}\varepsilon _{ab\langle k}\biggl[\!\frac{\left(
2\lambda \!-\!\rho \!-\!3p\right) }{3}D^{a}\langle \widehat{\sigma }%
_{j\rangle }^{b}\rangle \!-\!\langle \widehat{K}\!-\!\frac{\widehat{\Theta }%
}{3}\rangle D^{a}\pi _{j\rangle }^{b}\!-\!D^{a}\pi _{j\rangle c}\langle 
\widehat{\sigma }^{cb}\rangle \!-D^{a}\langle \widehat{\sigma }_{j\rangle
c}\rangle \pi ^{cb}\! \\
\!\!\!-\pi _{j\rangle c}D^{a}\langle \widehat{\sigma }^{cb}\rangle -\langle 
\widehat{\sigma }_{j\rangle c}\rangle D^{a}\pi ^{cb}-q_{j\rangle
}D^{a}\langle \widehat{K}^{b}\rangle \!-\!\langle \widehat{K}_{j\rangle
}\rangle D^{a}q^{b}\!-q^{b}D^{a}\langle \widehat{K}_{j\rangle }\rangle
\!-\!\langle \widehat{K}^{b}\rangle D^{a}q_{j\rangle }\! \\
-\frac{2}{3}D^{a}\Delta \widetilde{\pi }_{j\rangle }^{b}\biggr]+\widetilde{%
\kappa }^{2}\omega _{\langle k}\biggl[\!3\pi _{j\rangle }^{a}\langle 
\widehat{K}_{a}\rangle \!-\!3\langle \widehat{\sigma }_{j\rangle
}^{a}\rangle q_{a}\!\!-\!2\left( \lambda \!+\!\rho \right) \langle \widehat{K%
}_{j\rangle }\rangle \!+\!2\langle \widehat{\Theta }\rangle q_{j\rangle
}-\!2\!\Delta \widetilde{q}_{j\rangle }\!\biggr]\ .
\end{gather*}

\section{Concluding remarks}

Both the average and the difference equations reduce to the corresponding
equations given in Subsection IV.D of Ref \cite{3+1+1}, by taking into
account that in the particular case of a symmetric embedding for quantities
defined with an odd (even) number of $n^{a}$, the conditions $\Delta
f=2f,\langle f\rangle =0$ ($\Delta f=0,\langle f\rangle =f$) hold. In
particular, the extrinsic curvature components $\mathcal{K}$ belong to the
first group.

\textit{Acknowledgements}: This work was supported by the Pol\'{a}nyi and
Sun Programs of the Hungarian National Office for Research and Technology
(NKTH), and by the Hungarian Scientific Research Fund (OTKA) grant 69036. We
acknowledge financial support from the organizers of the Grassmannian
Conference in Fundamental Cosmology (Grasscosmofun'09).

\end{document}